\documentstyle[12pt,epsf]{article}

\def\be{\begin{equation}}
\def\ee{\end{equation}}
\def\bea{\begin{eqnarray}}
\def\eea{\end{eqnarray}}

\begin{document}

\begin{flushright}
%%%IPM/P-99/??  \\
hep-ph/9905484
\end{flushright}

\pagestyle{plain} 

\def\e{{\rm e}}
\def\cs{\frac{1}{4(2\pi\alpha')^2}}
\def\CV{{\cal{V}}}
\def\haf{{\frac{1}{2}}}
\def\tr{{\rm Tr}}
\def\CV{{\cal{V}}}
\def\goes{\rightarrow}
\def\goal{\alpha'\rightarrow 0}

\begin{center}
{\Large {\bf Do Quarks Obey D-Brane Dynamics? II}}
\end{center}

\vspace{.15cm}

\begin{center}
Amir H. Fatollahi

\vspace{.3 cm}

{\it Institute for Studies in Theoretical Physics and Mathematics (IPM),}\\ 
{\it P.O.Box 19395-5531, Tehran, Iran,}\\

\vskip .1 in 

{\sl fath@theory.ipm.ac.ir}

\end{center} 
\vskip .02 in 
\begin{abstract}
The cross section for two bosonic D0-branes is calculated in limit 
$\alpha'\rightarrow 0$. It is found that the  cross section shows 
Regge behavior.
\end{abstract}

%%%%%%%%%%%%%%%%%%%%%%%%%%%%%%%%%%%%%%%%%%%%%%%%%%%%%%%%%%%%%%%%%%%%%%%%%%%%
\section{Introduction}

The idea of string theoretic description of gauge theories is an old idea 
\cite{Po3} \cite{Polya}. Despite of the years that passed on this idea, it
is also activating different researches in theoretical physics 
 \cite{Polyakov}\cite{Pes}\cite{GKP}\cite{mvpol}. 

On the other hand, in the last few years our understanding about string theory 
is changed dramatically; a stream which is called usually "second string
revolution" \cite{vafa}. The scope of this stream is presentation of
a unified string theory as a fundamental theory of the known interactions.
One of the most applicable tools in the above program are D$p$-branes \cite
{Po1,Po2}. It is conjectured, and confirmed by various tests, 
that these objects can be considered as a perturbative representation of 
nonperturbative charged solutions of the low energy of string theories.

It has been known for long time that hadron-hadron scattering processes
have two different behaviors depending on the amount of the momentum 
transfer \cite{Close}. At large momentum transfer interactions appear
as interactions between the hadron constituents, partons or quarks, 
and some qualitative similarities to electron-hadron scattering emerge.
At high energies and small momentum transfer Regge trajectories are 
exchanged, the same which was the first motivation for the stringy 
picture of strong interaction. Besides the good fitting between Regge 
trajectories and the mass of strong bound-states has 
remained unexplained yet \cite{Po3,silas} .

Deducing the above observations from a unified picture is the challenge
of theoretical physics and it is tempting to search for the application 
of the recent string theoretic progresses in this area. In this way one may 
find D$p$-branes good tools to take their dynamics as a proper effective 
theory for the bound-states of quarks and QCD-strings (QCD Electric Fluxes)
between them. The conventional string theory is believed to have relevant 
physics at Planck scale and beyond. To use the string theory tools for 
QCD-strings one should replace the parameters with those which are
important in QCD. The case is somehow similar to the variation from 
early days of string theory as the theory of strong interaction to string 
theory as the theory of gravity. Pushing this idea in \cite{9902} the 
potential between two D0-branes at rest was calculated 
and the result appeared in good agreement with those come from 
phenomenology for quarks \cite{lucha}. 
Here we concern the scattering problem. 
Based on the results of \cite{Fat} 
we calculate the  cross section for two bosonic D0-branes. 
It is found that 
the  cross section shows the Regge behavior. This behavior 
has been used some years ago to fit the hadron-hadron 
total cross section data successfully \cite{DL,HERA} (see also \cite{222,344}
for other recent application of this behavior). 
Also the obtained cross section exhibits a rich 
Polology which can be corresponded with Regge poles.

%%%%%%%%%%%%%%%%%%%%%%%%%%%%%%%%%%%%%%%%%%%%%%%%%%%%%%%%%%%%%%%%%%%%%%%%%%%
\section{On D-Branes}

D$p$-branes are $p$ dimensional objects which are defined as
(hyper)surfaces which can trap the ends of strings \cite{Po2}. 
One of the most interesting aspects of D-brane dynamics appear in their
{\it coincident limit}. In the case of coinciding $N$ D$p$-branes in a
(super)string theory, their dynamics are captured by a 
dimensionally reduced $U(N)$ (S)YM theory to 
$p+1$ dimensions of D$p$-brane world-volume \cite{W,Po2,Tay}. 

In case of D0-branes $p=0$, the above dynamics reduces to quantum mechanics
of matrices, because only time exists in the world-line. The bosonic part
of the corresponding Lagrangian is \cite{KPSDF} 
\footnote{
Ignoring the fermionic part is reasonable for phenomenological
considerations because of the absence of supersymmetry in present nature. }
\bea\label{1.1}
L=m_0  \tr\; \biggl(\haf  D_tX_i^2 +\cs[X_i,X_j]^2\biggl) , 
i,j=1,\cdot\cdot\cdot,d, 
\eea 
where $\frac{1}{2\pi\alpha'}$ and $m_0=(\sqrt{\alpha'}g_s)^{-1}$ 
are the string tension and the mass of D0-branes respectively. Here 
$D_t=\partial_t-iA_0$ acts as covariant derivative in the 0+1 dimensional
gauge theory. For $N$ D0-branes $X$'s are in adjoint representation of 
$U(N)$ and have the usual expansion $X_i=x_{ia} T_a$, 
$a=1,\cdot\cdot\cdot N^2$. 

In fact (\ref{1.1}) is the result of the truncation of the string theory 
calculations in the so-called {\it "gauge theory limit"} defined by
\bea\label{1.10}
g_s &\rightarrow& 0,\nonumber\\
l_s &\rightarrow& 0,\nonumber\\
\frac{v}{l_s^2}&=&{\rm fixed},\nonumber\\
\frac{b}{l_s^2}&=&{\rm fixed},
\eea
which $v$ and $b$ are the velocities and distances relevant to 
the problem and $l_s=\sqrt{\alpha'}$ is the string length.

Firstly let us search for D0-branes in the above Lagrangian:\\ 
For each direction $i$ there are $N^2$ variables and not $N$ which one
expects for $N$ particles. Although there is 
an ansatz for the equations of motion
which restricts the $U(N)$ basis to its $N$ dimensional Cartan
subalgebra. This ansatz causes vanishing the potential and one
finds the equations of motion for $N$ free particles. In this case the $U(N)$
symmetry is broken to $U(1)^N$ and the interpretation of $N$ remaining
variables as the classical (relative) positions of $N$ particles is
meaningful. The center of mass of D0-branes is represented by the trace
of the $X$ matrices.

In the case of unbroken gauge symmetry, the $N^2-N$ non-Cartan elements have
a stringy interpretation, governing the dynamics of low lying 
oscillations of strings stretched
between D0-branes. Although the gauge transformations mix the entries
of matrices and the interpretation of positions 
for D0-branes remains obscure \cite{Ba},
 but even in this case the center of mass is meaningful. So the
ambiguity about positions only comes back 
to the relative positions of D0-branes.

Let us concentrate on the limit $\goal$. 
In this limit to have a finite energy one has 
\bea\label{1.15}
[X_i,X_j]=0,\;\;\;\forall\;i,j,
\eea
and consequently vanishing the potential term in the action.
So D0-branes do not interact and the action reduces to the
action of $N$ free particles 
\bea\label{1.20}
S=\int dt \sum_{a=1}^N \haf m_0 \dot{x}_a^2.
\eea
But the above observation fails in the times which D0-branes arrive
each other. When two D0-branes come very near each other two eigenvalues
of $X_i$ matrices will become approximately equal and this 
make the possibility that the 
corresponding off-diagonal elements take non-zero values. 
In fact this is the same story of gauge symmetry restoration.
In summary one may deduce that in the limit $\goal$ D0-branes
do not interact with each other except for when they coincide.

In the coincident limit the dynamics is complicated. 
The matrix position may be taken as:
\bea
X_i= \left( \matrix{x_i & Y_i \cr
Y_i^* & -x_i }\right),
\eea
where $Y^*$ is the complex conjugate of $Y$. By inserting this matrix in 
the Lagrangian one obtains:
\bea
S=\int dt\haf \bigg( (2m_0) \dot{X_i}^2 +  m_0 \dot{Y}_i\dot{Y}_i^* 
-  m_0 \cs (1-\cos^2\theta) 
x_j^2 Y_iY_i^*+ m_0 \dot{x}_j^2+O(Y^3)\bigg),
\eea
with $X$ for the center of mass and $\theta$ is the angle between $\vec{x}$ 
and the complex vector $\vec{Y}$. 
As is apparent in the limit $\goal$ which is in our direct interest 
the $x$ element can not take large values and have a small range of 
variation. In high-tension approximation of strings, one takes the 
relative distance of D0-branes constant of order $x\sim g_s^{1/3}l_s$
\footnote{This length is the size of the D-particle bound-states \cite{Fat}}. 
So one writes:
\bea
S=\int dt \bigg(\haf (2m_0) \dot{X_i}^2 + 
\haf m_0 \dot{Y}_\perp\cdot\dot{Y}_\perp^*
- \haf m_0 \frac{k^2g_s^{2/3}}{\alpha'}  
Y_\perp\cdot Y_\perp^* +\cdot\cdot\cdot
\bigg),
\eea
where in the above $k$ is an $\alpha'$ independent numerical factor, and 
$Y_\perp$ is the perpendicular part of the $\vec{Y}$ to the relative distance 
$\vec{x}$. The parallel part of $\vec{Y}$ behaves as a free part.
In $d+1$ dimensions of space-time the dimension of $Y_\perp$ is 
$d-1$ which shows that we are encountered with $2\times(d-1)$ 
harmonic oscillators because, $Y$ is a complex variable
\footnote{This is the same number of harmonic oscillators which 
appear in one-loop calculations \cite{9902}.}. These harmonic oscillators 
are corresponded to vibrations of oriented open strings stretched between 
D0-branes.
%%%%%%%%%%%%%%%%%%%%%%%%%%%%%%%%%%%%%%%%%%%%%%%%%%%%%%%%%%%%%%%%%%%%%%%%%%
%%%%%%%%%%%%%%%%%%%%%%%%%%%%%%%%%%%%%%%%%%%%%%%%%%%%%%%%%%%%%%%%%%%%%%%%%%
\section{Scattering Amplitude}

Substructure of hadrons are probed in a sufficiently 
large momentum transfer scattering processes of a fundamental 
particle, e.g. an electron. The existence of a point-like substructure, 
"parton", is the result of the {\it "scaling"} behavior of some special 
functions, i.e. the absence of any {\it "scale"} is related to point-like
objects.

One of the ingredients of the above picture
is the assumption of free partons in the suddenly collision 
processes \cite{Close}. Accordingly it is assumed that 
in a collision process the electron sees a free parton 
instead of the hadron as a whole.

With the above in mind it is reasonable to calculate the scattering 
amplitude between two individual D0-branes to find a flavor
about the behavior of the scattering amplitude of two hadrons
which D0-branes are assumed as their partons. Also it is natural 
to assume that this result is for elastic-large momentum transfer 
(Deep Elastic) regime of hadron collisions.

Here we use the result of \cite{Fat}. In \cite{Fat} it is shown 
that the quantum traveling of D0-branes can be corresponded with field
theory Feynman graphs and their associated amplitudes in the light-cone 
frame. In the following we review the approach to calculate the 
amplitude.

For two D0-branes take the probability amplitude presented by 
path integral as
\bea\label{1.25}
\langle x_3,x_4;t_f| x_1, x_2;t_i  \rangle=\int \e^{-S}.
\eea
In the limit $\goal$ in that parts of paths which D0-branes 
are not identified, only the diagonal matrices have contribution
to the path integral. This is because of large value of 
action in the exponential. So the action in 
the path integral reduces to the action of two free 
D0-branes for non-coincident parts of paths, e.g. 
$(x_1,x_2)$ till $X_1$ in the Fig.1.
Accordingly one may write, Fig.1 
\footnote{Here as the same which one does in field theory we
have dropped the dis-connected graphs.},
\bea\label{1.30}
&~&\langle x_3,x_4;t_f| x_1, x_2;t_i  \rangle=\left[\int \e^{-S}\right]_{\goal}=
  \int_{t_i}^{t_f} (dT_1  dT_2 )
  \int_{-\infty}^{\infty} (dX_1 dX_2 ) \nonumber\\
&~&\times  
\bigg(K_{m_0}(X_1,T_1;x_1,t_i) K_{m_0}(X_1,T_1;x_2,t_i) \bigg)    \nonumber\\
&~&\times
\bigg(K_{2m_0}(X_2,T_2;X_1,T_1)
K_{oscillator}(Y_\perp=0,T_2;Y_\perp=0,T_1)\bigg)
\nonumber\\
&~&\times
 \bigg(K_{m_0}(x_3,t_f;X_2,T_2) K_{m_0}(x_4,t_f;X_2,T_2)\bigg),
\eea
which $K_m(y_2,t_2;y_1,t_1)$ is the non-relativistic propagator 
of a free particle with mass $m$ between $(y_1,t_1)$ and $(y_2,t_2)$ and
$K_{oscilator}(Y_\perp=0,T_2;Y_\perp=0,T_1)$ is the harmonic oscillator 
propagator. $\int dT_1 dT_2 dX_1 dX_2$ is for a summation over 
different "Joining-Splitting" times and points. We use 
in $d$ dimensions the representations
$$
K_{m}(y_2,t_2;y_1,t_1)=\theta(t_2-t_1)\frac{1}{(2\pi)^d} \int d^d p 
\;\e^{ip\cdot(y_2-y_1)-\frac{ip^2(t_2-t_1)}{2m}},
$$
$$
K_{oscilator}(Y_\perp=0,T_2;Y_\perp=0,T_1)=
\theta(T_2-T_1)\bigg(\frac{m_0\omega}{2\pi i 
\sin[\omega(T_2-T_1)]}\bigg)^{d-1},
$$ 
where $\theta(t_2-t_1)$ is the step function and 
$\omega$ is the harmonic oscillator frequency here to be 
$kg_s^{1/3}/l_s$. 
Because of complex nature of $Y_\perp$ the
power for the harmonic propagator is twice of $\frac{d-1}{2}$.

%%%%%%%%%%%%%%%%%%%%%%%%%%%%%%%%%%%%%%%%%%%%%%%%%%%%%%%%%%%%%%%%
\begin{figure}[t]
\begin{center}
\leavevmode
\epsfxsize=80mm 
\epsfysize=80mm
\epsfbox{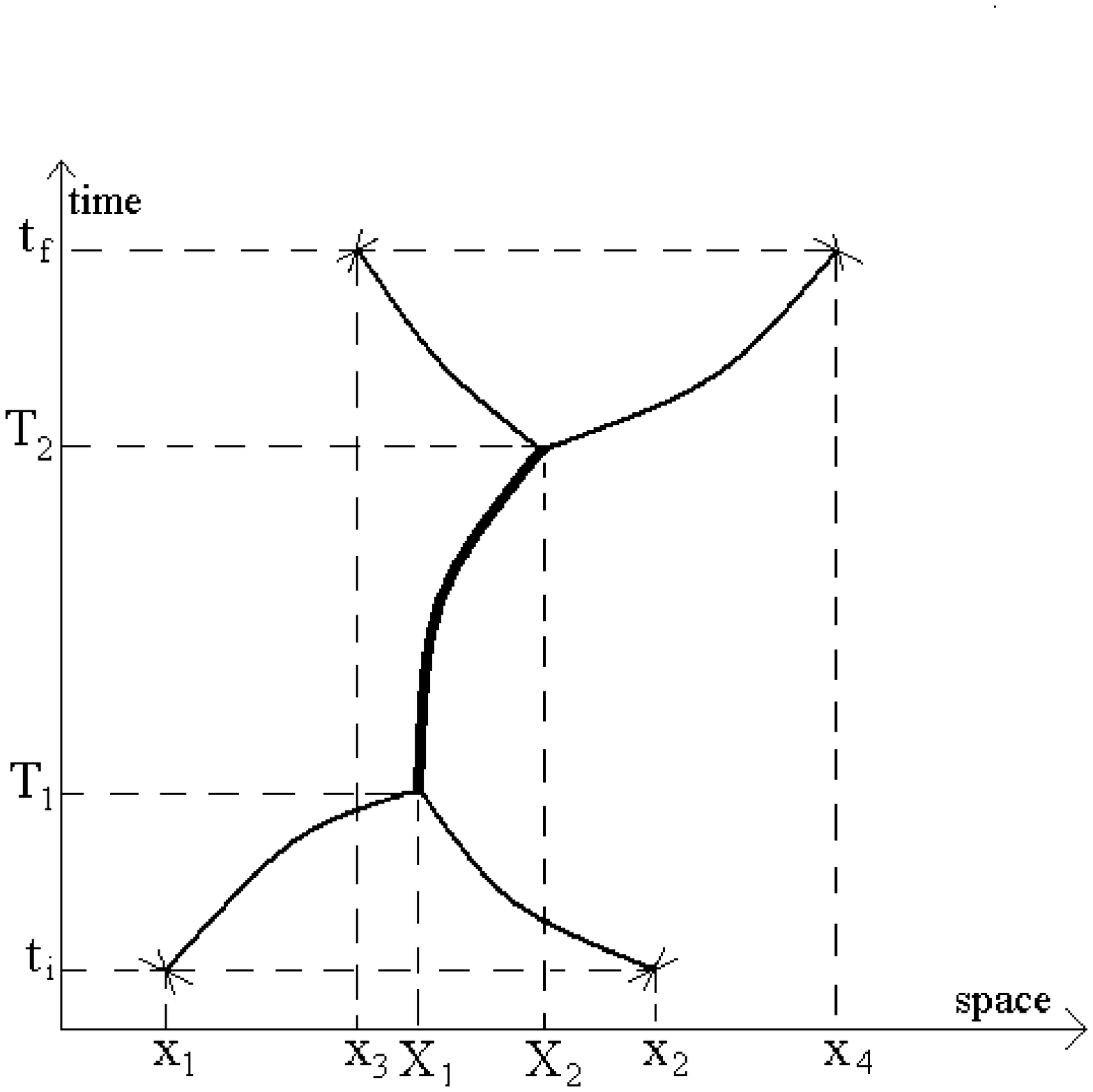}
\caption{{\it A typical tree path for D0-branes}}
%\label{fig:?}
\end{center}
\end{figure}

%\begin{center}
%\begin{figure}[t]
%\end{picture}
%\vspace{-1cm}
%\caption{{\it A typical path.}}
%\end{figure}
%\end{center}
%%%%%%%%%%%%%%%%%%%%%%%%%%%%%%%%%%%%%%%%%%%%%%%%%%%%%%%%%%%%%%%%

Translating all the above to the momentum space is obtained by
($E_k=\frac{p_k^2}{2m_0}$ with $k=1,2,3,4$)
\bea\label{1.32}
\langle p_3,p_4;t_f |p_1, p_2;t_i  \rangle\sim \e^{i(E_3+E_4)t_f-i(E_1+E_2)t_i}
&~&\int \prod_{a=1}^4 dx_a \e^{i(p_1x_1+p_2x_2-p_3x_3-p_4x_4)}
\nonumber\\&~&\times\langle x_3,x_4;t_f | x_1, x_2;t_i \rangle.
\eea
This representation is useful to calculate the cross section.
The integrals can be done easily to find
\bea
&~&\langle p_3,p_4;t_f |p_1, p_2;t_i\rangle            \sim
\delta^d(p_1+p_2-p_3-p_4)
\int_{t_i}^{t_f} dT_1 dT_2 \theta(T_2-T_1)
\nonumber\\
&~&\times\exp(\frac{-i(p_1^2+p_2^2)T_1}{2m_0})
\exp(\frac{-iq^2(T_2-T_1)}{4m_0})
\exp(\frac{i(p_3^2+p_4^2)T_2}{2m_0})
\nonumber\\
&~&
K_{oscillator}(Y_\perp=0,T_2;Y_\perp=0,T_1)
\eea
where $\vec{q}=\vec{p}_1+\vec{p}_2=\vec{p}_3+\vec{p}_4$.

To have a real scattering process one assumes
$$
t_i\rightarrow -\infty,\;\;\;t_f\rightarrow \infty.
$$
We put $T\equiv T_2-T_1$ which has the range $0\leq T \leq \infty $. 
The integrals yield
\bea
\langle p_3,p_4;\infty |p_1, p_2;-\infty\rangle &\sim&
\delta^d(p_1+p_2-p_3-p_4)
\delta(\frac{p_1^2}{2m_0}+\frac{p_2^2}{2m_0}-
\frac{p_3^2}{2m_0}-\frac{p_4^2}{2m_0})
\nonumber\\
&~&
\int_0^\infty dT\; \e^{\frac{-iT}{4m_0}(q^2-2(p_1^2+p_2^2))}
\bigg(\frac{m_0\omega}{\sin (\omega T)}\bigg)^{d-1},
\eea
By recalling the the energy-momentum relation in Light-Cone gauge \cite{Fat}
one has:
$$
2(p_1^2+p_2^2)-q^2= 2(2m_0)(\frac{p_1^2+p_2^2}{2m_0})-q^2=
2q_+q_--q^2=q_\mu q^\mu\equiv q_\mu^2.
$$
So it is found:
\bea
\langle p_3,p_4,E_3,E_4;\infty |p_1,p_2,E_1,E_2;-\infty\rangle 
&\sim& \delta^d(p_1+p_2-p_3-p_4)
\delta(E_1+E_2-E_3-E_4)\nonumber\\&~&
\int_0^\infty dT \e^{\frac{-q_\mu^2}{4m_0}T}
\bigg(\frac{m_0\omega}{\sin (\omega T)}\bigg)^{d-1}.
\eea
We perform a cut-off for $T$ in small values as 
$0 < \epsilon\leq T \leq \infty $, 
with $\epsilon$ be small. By changing the integral 
variables as $\e^{-2\omega T}=\eta$ we have
\bea\label{amp}
\langle p_3^\mu,p_4^\mu;\infty |p_1^\mu,p_2,^\mu;-\infty\rangle &\sim&
\delta^d(p_1+p_2-p_3-p_4)
\delta(p_{-1}+p_{-2}-p_{-3}-p_{-4})\nonumber\\&~&
\frac{(m_0\omega)^{d-1}}{2\omega}
\int_0^x d\eta\; \eta^{\frac{-q_\mu^2}{8m_0\omega}+\frac{d-3}{2}}
(1-\eta)^{-d+1},\nonumber\\
&\sim&
\delta^d(p_1+p_2-p_3-p_4) \delta(p_{-1}+p_{-2}-p_{-3}-p_{-4})
\nonumber\\&~&
\frac{(m_0\omega)^{d-1}}{2\omega}
B_x(\frac{-q_\mu^2}{8m_0\omega}+\frac{d-1}{2},-d+2)
\eea
with $1\sim x=\e^{-2\omega\epsilon}$ and $B_x$ is the Incomplete 
Beta function. The longitudinal momentum conservation trivially 
is satisfied. Besides because of conservation of this momentum one 
can not expect so-called $t$-channel processes.

%%%%%%%%%%%%%%%%%%%%%%%%%%%%%%%%%%%%%%%%%%%%
\vspace{.5cm}
{\bf Polology}

Equivalently one may use the other representation of $K_{oscillator}$
as
\bea
K_{oscillator}(Y_\perp=0,T_2;Y_\perp=0,T_1)=
\sum_{n} \langle0|n\rangle\langle n|0\rangle \e^{-iE_n(T_2-T_1)},
\eea 
with $E_n$'s as the known $H_{oscillator}$ eigenvalues. By this 
representation one finds the pole expansion \cite{Fat}:
\bea
\langle p_3,p_4;\infty |p_1, p_2;-\infty  \rangle&\sim&
\delta^d(p_1+p_2-p_3-p_4)
\delta(p_{1+}+p_{2+}-p_{3+}-p_{4+})
\nonumber\\
&~&\times \lim_{\epsilon \rightarrow 0^+}
\sum_n C_n\; \frac{i4m_0}{ q_\mu q^\mu - M_n^2+i\epsilon}.
\eea

This pole expansion also can be derived by extracting the poles  
of the amplitude (\ref{amp}) with the condition
\bea
\frac{-q_\mu^2}{8m_0\omega}+\frac{d-1}{2}=-n, \;\;
{\rm with}\; n\; {\rm as\; a\; positive\; integer},
\eea
or
\bea
M_n^2=\frac{8k(n+\frac{d-1}{2})}{g_s^{2/3}l_s^2}.
\eea

%%%%%%%%%%%%%%%%%%%%%%%%%%%%%%%%%%%%%%%%%%%%%%%%%%%%%%%%%%%%%%%%%%%%%%%%%%%%%%
\section{Conclusion and Discussion}

In this letter we calculate the cross section of two bosonic D0-branes 
to find a flavor about the cross section behavior of two hadrons with 
D0-branes as their partons. It is natural to take the results in 
elastic-large momentum transfer (Deep Elastic) regime. It was found that
the cross section shows Regge behavior; the behavior with a long sounds
in theoretical physics. This behavior has been used to insight to some 
aspects of hadron physics \cite{DL,222,344}.

{\it Why non-commutativity?}\\
Special relativity in a modern compact definition may be 
represented as follows:\\
{\sl A modification of space-time to prepare it  
as a ground for the natural and theoretically consistent 
propagation of fields.}

So one learns that the space-time makes a $X_\mu$ 4-vector which behaves
like the electromagnetic gauge field $A_\mu$ (spin 1) under the 
boost transformations. 

Also in this way supersymmetry (SUSY) 
is a natural continuation of the special relativity 
program:
\\Including spin $\frac{1}{2}$ sectors to the coordinates of space-time,
as the fermions of nature.
This leads one to the space-time formulation of the SUSY theories. 
Also it is the same way which one introduces fermions to the bosonic
string theory.

Now, what may be modified if nature has non-abelian (non-commutative) 
gauge fields? In present nature non-abelian gauge fields can not 
make spatially long coherent states; they are confined or too heavy. 
But the picture may be changed inside a hadron or very near of an 
electron. In fact recent developments of string theories sound this 
change and it is understood that non-commutative coordinates and
non-abelian gauge fields are two sides of one coin.
The future theoretical research in this area may make clear the
relations.
%%%%%%%%%%%%%%%%%%%%%%%%%%%%%%%%%%%%%%%%%%%%%%%%%%%%%%%%%%%%%%%%%%%%%%%%%%%%%%

\vspace{.2cm} 
{\bf Acknowledgement}

I am grateful to S. Parvizi for useful discussions. 
%%%%%%%%%%%%%%%%%%%%%%%%%%%%%%%%%%%%%%%%%%%%%%%%%%%%%%%%%%%%%%%%%%%%%%%%%%%%%%%%%%%%%%%%%
%%%%%%%%%%%%%%%%%%%%%%%%%%%%%%%%%%%%%%%%%%%%%%%%%%%%%%%%%%%%%%%%%%%%%%%%%%%%%%%%%%%%%%%%%%%%%%%%%%%%%%%%%%%%%%%%%%%%%%%%%%%%%%%%%%%%%%%%%%%%%%%%%%%%%%%%%%%%%%%%%%%%%%%%%%%
%%%%%%%%%%%%%%%%%%%%%%%%%%%%%%%%%%%%%%%%%%%%%%%%%%%%%%%%%%%%%%%%%%%%%%%%%%%%%%%%%%%%%%%%%%%%%%%%%%%%%%%%%%%%%%%%%%%%%%%%%%%%%%%%%%%%%%%%%%%%%%%%%%%%%%%%%%%%%%%%%%%%%%%%%%%

\end{document}